\documentclass[aps,prb,twocolumn,superscriptaddress,floatfix,eqsecnum, showpacs]{revtex4}
\usepackage{amsmath}
\usepackage{amssymb}
\usepackage{wrapfig}
\usepackage{subfigure}
\usepackage{psfrag}
\usepackage{tabularx}

\usepackage[dvips]{graphicx}
\DeclareGraphicsExtensions{.eps}

\begin{document}
\title{Competition between Pomeranchuk instabilities in the nematic and hexatic channels in a two-dimensional spinless Fermi fluid}

\author{Daniel G. Barci}
\affiliation{Departamento de F{\'\i}sica Te\'orica Universidade do
Estado do Rio de Janeiro. Rua S\~ao Francisco Xavier 524, 20550-013,
Rio de Janeiro, RJ, Brazil.}
\altaffiliation{Research Associate of the Abdus Salam International Centre for Theoretical Physics, ICTP, Trieste, Italy}

\author{Marta Trobo}
\affiliation{Instituto de F\'{\i}sica de La Plata, CONICET,
Argentina} \affiliation{Departamento de F\'{\i}sica, Universidad
Nacional de La Plata, Argentina.}

\author{Victoria Fern\'andez}
\affiliation{Instituto de F\'{\i}sica de La Plata, CONICET,
Argentina} \affiliation{Departamento de F\'{\i}sica, Universidad
Nacional de La Plata, Argentina.}

\author{Luis E. Oxman}
\affiliation{Instituto de F\'{\i}sica, Universidade Federal
Fluminense, Campus da Praia Vermelha, Niter\'oi, 24210-340, RJ,
Brazil. }

\date{\today}

\begin{abstract}
We study the competition between the nematic and the hexatic phases of a  two-dimensional spinless Fermi fluid near Pomeranchuk instabilities.
We show that the general phase diagram of this theory contains a bicritical point where  two second order lines and a first order nematic/hexatic phase transition meet together. We found that at criticality, and deep inside the associated symmetry broken phases, the low energy theory is governed by a dissipative cubic mode, even near the bicritical point where nematic and hexatic fluctuations cannot be distinguished due to very strong dynamical couplings.  
\end{abstract}


\maketitle

\section{Introduction}

Nowadays, there is a large amount of theoretical and experimental work, studying new phases of 
strongly correlated fermionic systems that spontaneously break rotational and/or translational symmetry. 
These electronic states were called {\em quantum liquids crystals}\cite{Nature} because they are anisotropic metals sharing the same symmetry 
properties as usual liquid crystals. Today, we have several examples of 
smectic\cite{smectic} and nematic\cite{Vadim} quantum liquid phases. 
It is interesting that, at mesoscopic scale, we can understand the basic physics of these phases just on symmetry grounds, without relying on the detailed microscopic description. Of course, the final fate of the phase depends on thermal and/or quantum  fluctuations, and on dimensionality.  

The quantum nematic state is probably the best candidate\cite{Manousakis} to explain the anisotropies observed in 2DEG at half filled Landau levels\cite{QHSExp}. This state is also expected to appear in other strongly correlated systems like high $T_c$ superconductors\cite{Kivelson1,Kivelson2} and in heavy fermion compounds\cite{Grigera}. 

The first consistent description of the quantum nematic state was done in ref. \onlinecite{Vadim}, and its nonperturbative one-particle properties were studied in ref. \onlinecite{Lawler}.
Several important results were obtained in the case of the two-dimensional isotropic-nematic quantum phase transition\cite{Lawler,CastroNeto,Lawler2,Kim}. At criticality, the low energy properties are ruled by a dissipative collective cubic mode $\omega\sim i q^3$. The coupling of this mode with fermions wipes off the quasi-particle pole in the spectral functions (except for some symmetrical points), implying that the isotropic-nematic transition, from the electronic point of view, is a Fermi/non-Fermi liquid phase transition.  
These calculations, initially done in Hartree-Fock approximation\cite{Vadim}, were confirmed with a nonperturbative treatment of the Pomeranchuk instability, using multidimensional bosonization, \cite{Lawler}  and with the more usual Landau theory of Fermi liquids\cite{CastroNeto}. 

In recent years, there was an increasing interest on Pomeranchuk instabilities\cite{Pomeranchuk1} not only in continuous models but also in the lattice\cite{Vadim, BL, Lawler, Pomeranchuk2, Andy, Congjun, Wegner, Kee}.  Concerning quantum Hall samples at moderate magnetic fields, there is experimental evidence\cite{QHSExp,exp} and theoretical proposals\cite{F-K, wex} pointing to the idea that a huge number of phases are present, which depend on the filling factor and temperature, that is, a rich and delicate competition between several liquid states (metallic) and crystal states (insulators). 
Very near integer filling factor, a Wigner crystal state is by now well established\cite{WignerCrystal, Foglercrystal}. Thermal fluctuations of this state could melt the crystal into an hexatic phase, which is homogeneous and anisotropic, with the residual symmetry of the Wigner triangular lattice. Increasing fluctuations would produce more disorder, leading to an hexatic/isotropic transition. There is a clear region in the phase diagram where the isotropic, nematic and hexatic phases compete very closely\cite{F-K}.
The aim of this paper is to characterize this quantum phase transition as a first step to understand real Hall liquids. 

In order to simplify, we study a version of Landau theory for spinless Fermi liquids\cite{Pines}, where we consider the effect of the curvature of the fermionic dispersion relation\cite{BL} around the Fermi surface and possibly four-body interactions. In particular, we analyze  quantum fluctuations in a region where the Landau parameters $F_2$ and $F_6$ are very near the Pomeranchuk instability.  The main results are depicted in figure (\ref{figbp}),  where we draw a mean field phase diagram for the Fermi liquid in terms of the Landau parameters. We clearly see two second order lines, corresponding to the isotropic/nematic and the isotropic/hexatic phase transitions. These transitions meet together at a bi-critical point where a first order nematic/hexatic transition emerges. 

Along the two second order lines, the nematic and hexatic order parameters are weakly coupled and we expect a similar behavior of these two phases. However, very near the bi-critical point the coupling is very strong and it is not possible to distinguish between nematic and hexatic fluctuations. We find that the critical theory {\em at the bicritical point} is  governed by a low lying collective mode with dynamical exponent $z=3$.
We also found that, in the symmetry broken phase,  the orientations of the nematic and hexatic principal axes are not independent, they differ by specific angles dictated by symmetry.  In the two ordered phases, the Goldstone mode associated with the angle fluctuation is also a dissipative cubic mode, retaining the criticality of the theory deep inside the two symmetry broken phases.  Near the bicritical point, they are separated by a discontinuous transition  and the dynamics in this region is related with metastability. 

In the next sections we show the details of the model and the main reasoning leading to the above mentioned results.

 \section{The 2D Isotropic-Nematic-Hexatic phase transition \label{INH}}

\begin{figure}
\centering
\includegraphics[width= 0.40\textwidth]{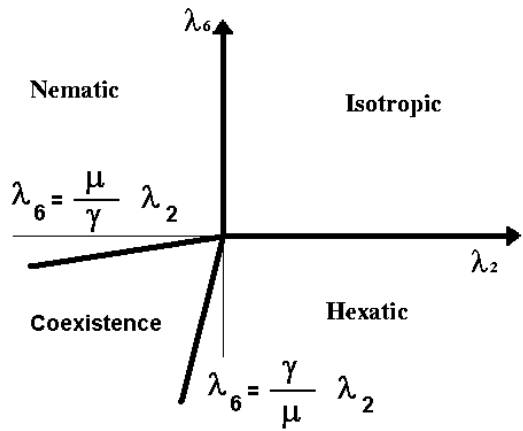}
\caption{Phase diagram for the free energy in eq. \ref{Fcomplex}, for $\gamma>\mu$. The bold lines represent second order phase transitions and, in the case of Fermi liquids, the control parameters are given by $\lambda_2=1+F_2$ and $\lambda_6=1+F_6$. For  $\beta=0$, the symmetry of the model is $U(1)\times U(1)$, thus allowing the coexisting phase. However, for $\beta\ne 0$ this phase turns out to be nematic.}
 \label{figtp}
\end{figure}

\begin{figure}
\centering
\includegraphics[width=.45\textwidth]{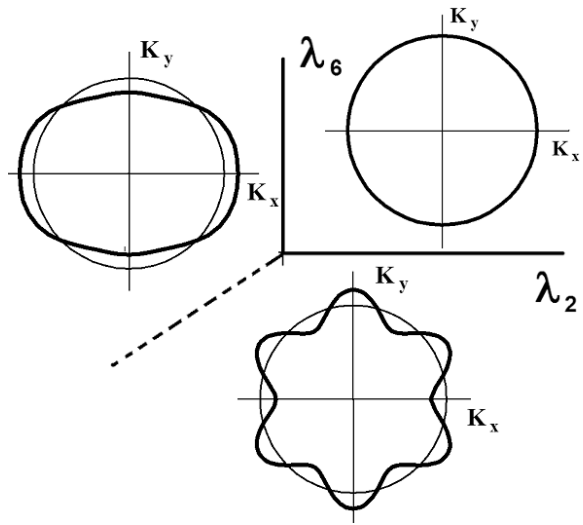}
\caption{Phase diagram for the free energy in eq. \ref{Fcomplex}, for $\gamma<\mu$. The bold lines represent second order phase transitions, while the dot line corresponds to a first order transition. In the insets, we draw the Fermi surfaces computed  with eq. (\ref{kF}). The relative angle between the orientation of the principal axis between the nematic and hexatic phases was fixed to $\pi/2$.}
  \label{figbp}
\end{figure}

The order parameters for two-dimensional nematic and hexatic phases can be cast in terms of complex fields, namely,
\begin{eqnarray}
\psi_2&=&\rho_2\; e^{i 2 \chi_2}\nonumber \\ 
\psi_6&=&\rho_6\; e^{i 6 \chi_6},
\label{orderparameter}
\end{eqnarray}
where $\psi_2$ is the nematic order parameter and $\psi_6$ is the hexatic one.  The complex representation is typical of two dimensions in which  the rotation group $O(2)$ is isomorphic to the unitary group $U(1)$. 
Defined in this way, the parameters  in eq. (\ref{orderparameter}) have nematic and hexatic symmetries $\chi_2\to\chi_2+\pi$ and $\chi_6\to\chi_6+\pi/3$, respectively. 

Near a phase transition the order parameters are small so that we can write down a polynomial free energy, keeping just quartic terms in the expansion. At mean field, considering uniform $\psi_2$ and $\psi_6$, the more general local and rotational invariant free energy is,
\begin{eqnarray}
F&=& \frac{\lambda_2}{2} \psi_2^*\psi_2+ \frac{\lambda_6}{2} \psi_6^*\psi_6+ \frac{\gamma}{4} \left( |\psi_2|^4+|\psi_6|^4 \right)
\nonumber \\
&+&\frac{\mu}{2} |\psi_2|^2|\psi_6|^2 
+\frac{\beta}{2}
\left(\psi_2^3 \psi_6^*+\mbox{c.c.}
\right),
\label{Fcomplex}
\end{eqnarray}
where ``c.c'' means complex conjugate.
In order to simplify, we have considered the same $\gamma>0$ for the quartic terms associated with both order parameters, however, we could allow different values  without changing any substantial physics. We also consider $\mu>0$, while $\lambda_2$ and $\lambda_6$ are control parameters of the phase transitions.
The parameter $\beta$ is special, since it changes the symmetry of the system. If $\beta=0$, the system is invariant under a global  $U(1)\times U(1)$ transformation, since we can change the phases of the two order parameters independently. However if $\beta\neq 0$ the symmetry is reduced from  $U(1)\times U(1)\to U(1)$. This is more evident if we rewrite eq. \ref{Fcomplex} in polar form, 
\begin{eqnarray}
F&=& \frac{\lambda_2}{2} \rho_2^2+ \frac{\lambda_6}{2} \rho_6^2+ \frac{\gamma}{4} \left(\rho_2^4+\rho_6^4 \right)
+\frac{\mu}{2} \rho_2^2\rho_6^2 \nonumber \\
&+&\beta
\rho_2^3\rho_6\cos 6(\chi_2-\chi_6)\; ;
\label{Freal}
\end{eqnarray}
the free energy depends on the difference between the two angles $\chi_2-\chi_6$, preserving global rotational invariance. 

To study the mean field phase diagram, we minimize the free energy with respect to the independent variables $\rho_2$, $\rho_6$ and $\chi_2-\chi_6$. The minimization with respect to the angles is straightforward. If $\beta>0$, $\chi_2-\chi_6$ will take the values $1/6\;\pi$, $1/2\;\pi$ or $5/6\;\pi$ to make the last term in eq. (\ref{Freal}) negative.   On the other hand, if $\beta<0$, $\chi_2-\chi_6$ will take the values $0$, $1/3\;\pi$ or $2/3\;\pi$. Therefore, the principal axes of the nematic and hexatic phases will not necessarily  be aligned. This will have very interesting consequences, specially in the case of first order transitions where we can have metastable liquid crystal states with different principal axes. 

After optimizing the angle variables, let us analyze the phase diagram associated with, 
\begin{equation}
F= \frac{\lambda_2}{2} \rho_2^2+ \frac{\lambda_6}{2} \rho_6^2+ \frac{\gamma}{4} \left(\rho_2^4+\rho_6^4 \right)+\frac{\mu}{2} \rho_2^2\rho_6^2-\beta
\rho_2^3\rho_6,
\label{Frho}
\end{equation}
considering all coefficients positive except, of course, our control parameters $\lambda_2$ and $\lambda_6$. 

We begin by considering the case $\beta=0$. In this case, the free energy is very similar to that occurring in several antiferromagnetic systems with weak anisotropy\cite{Lubensky} and presents in general multicritical points.  The results are summarized in figs. \ref{figtp} and \ref{figbp}. Note that $\rho_2=\rho_6=0$ are extrema of the free energy for any value of the parameters. In particular, if 
$\lambda_2>0$ and $\lambda_6>0$ they are the absolute minima, representing the isotropic phase. 
When $\lambda_2,\lambda_6$ switch to negative values, different types of solutions emerge, depending on the other parameters of the model. 
For instance, for $\lambda_6<0$, and for any value of $\lambda_2$, we find the solution $\rho_2=0$, $\rho_6^2=-\lambda_6/\gamma$. On the other hand, if $\lambda_2<0$, for any value of $\lambda_6$, the solution is  $\rho_2^2=-\lambda_2/\gamma$, $\rho_6=0$. 
These solutions determine two  second order phase transitions: isotropic/nematic and isotropic/hexatic, in the $\lambda_2=0$, $\lambda_6>0$ and $\lambda_2>0$, $\lambda_6=0$ semi-axis, respectively, see figs. \ref{figtp} and \ref{figbp}.

If $\lambda_2$ and $\lambda_6$ are negative, there are in addition new solutions where both order parameters are different from zero, 
\begin{eqnarray}
\rho_2^2&=&\frac{-\gamma\lambda_2+\mu\lambda_6}{\gamma^2-\mu^2} \label{mix1} ,\\
\rho_6^2&=&\frac{\mu\lambda_2-\gamma\lambda_6}{\gamma^2-\mu^2}. \label{mix2}
\end{eqnarray}
Which of all these extrema are  absolute minima depends essentially on the relative values of $\gamma$ and $\mu$. Indeed, 
if $\gamma^2>\mu^2$,  the solutions of eqs. (\ref{mix1}) and (\ref{mix2})  are valid in the region $\gamma/\mu\;\lambda_2<\lambda_6 <\mu/\gamma\;\lambda_2$.
Therefore, the straight lines $\lambda_6=\gamma/\mu\;\lambda_2$ and $\lambda_6 =\mu/\gamma\;\lambda_2$ determine the two second order phase transitions shown in fig. \ref{figtp}.  Inside this region, the pure nematic and the pure hexatic solutions become unstable, leading to a coexistence phase, where the two order parameters are different from zero.  This region is possible due to the $U(1)\times U(1)$ symmetry, which makes the phases of the two order parameters independent. 
Notice that the region of coexistence is controlled by the coupling $\mu$ between the two order parameters. If $\mu\to 0$, they decouple and this region covers the complete third quadrant, as it should be. 
For higher $\mu$, the region is stretched between the two lines shown in fig. \ref{figtp}, and for the limiting value $\mu=\gamma$, the area of coexistence shrinks to zero, and the second order transition collapses into a first order one, valid in the case of $\mu>\gamma$, this is depicted in fig. \ref{figbp}.

If $\gamma^2<\mu^2$, the solutions of eqs. (\ref{mix1}) and (\ref{mix2}) (valid now in the region  $\mu/\gamma\;\lambda_2<\lambda_6 <\gamma/\mu<\lambda_2$) are unstable, and the true ground state is the pure nematic   and the pure hexatic solution, separated by a first order line (see  fig. \ref{figbp}).  

Now, let us analyze the $\beta\ne 0$ case. If $\beta$ is large enough, the system will present unstable directions in the $(\rho_2,\rho_6)$ plane,  invalidating the quartic expansion of the free energy. Moreover, it is not difficult to realize that there is a critical value $\beta_c$ bellow which the theory is stable, in the case of having $\mu\sim\gamma$, $\beta_c\sim 0.77\;\gamma$.
As we mentioned, an important effect of the $\beta$ term is to reduce the symmetry of the model from $U(1)\times U(1)\to U(1)$ establishing a relationship between the relative phases of the two order parameters; a direct consequence is that the coexisting phase of fig. \ref{figtp} is now a nematic one,  since we cannot freely change the phases independently. 
On the other hand, the term $\beta<\beta_c$ does not change the character of the phase transition we have described. Therefore, in  fig. \ref{figtp}, the lines remain as second order transitions, except that we should consider now the coexisting phase as a nematic one. 

In the case  $\mu>\gamma$ (fig. \ref{figbp}), we will still have two second order lines, and a first order line between the nematic and hexatic phases. The explicit solutions can be evaluated pertubatively in $\beta$.
 
To be more precise, for the case $\mu>\gamma$, the isotropic and hexatic phases are the same as those we have described for $\beta=0$. However, when $\lambda_2<0$ and $\lambda_6>0$, $\rho_6$ assumes a value of order $\beta$ given by,
\begin{equation}
\rho_6=-\frac{\beta}{\gamma}\;\;\frac{|\lambda_2|^{3/2} \gamma^{1/2}}{\gamma |\lambda_6|+\mu|\lambda_2|}+ O((\beta/\gamma)^2).
\end{equation}
We are tempted to interpret this as a coexisting phase, however, as stated before,  we are not allowed to change the phases of the two order parameters independently. Then, the residual symmetry is the nematic one and the global phase is nematic with higher harmonics. This solution remains the global minimum deep inside the $\lambda_6<0$ region, when it finds the first order line depicted in fig. \ref{figbp}, and jumps discontinuously to the pure hexatic phase.
Therefore, the main effect of the $\beta$ term is to slightly modify the first order line $\lambda_2=\lambda_6$ and to add higher harmonics to the nematic solution. As we have discussed, this term is also responsible for the relative alignment between the principal axes of the nematic and hexatic phases. 

Of course, this analysis is only valid very near the multicritical point where the order parameters are small. Note that the different phases are separated by straight lines just because we are calculating at leading order in the order parameters. We expect that higher order corrections, as well as fluctuations, will curve this lines as occurs when studying multicritical behavior in anisotropic antiferromagnets\cite{Lubensky}. However, the character of the phase transitions will not change. 

An important point to discuss is whether this mean field phase diagram survives fluctuations or not. 
In the case of classical systems, thermal fluctuations will turn the second order phase transitions into Kosterlitz-Thouless\cite{KoTh1973}
type ones. The reason is that the angular correlations have logarithmic divergences that destroy the real order of the phase, keeping a quasi-long-range-order in the correlations\cite{Manousakis,ToNe1981}. We expect that the first-oder transition will be more robust against fluctuations, keeping its discontinuous character or possibly becoming slightly rounded.
  
At $T=0$, quantum fluctuations  depend on the dynamics of the Goldstone modes associated with the spontaneously broken symmetry. In the case of Fermi liquids, we will show in the next sections that fluctuations of the nematic  and hexatic order parameters provide a dynamical exponent $z=3$, implying an effective dimension equal to five, above the upper critical dimension. Therefore, the order of the transitions studied at mean field is expected to be valid in the quantum case.

\section{Pomeranchuk instabilities in the nematic and hexatic channels \label{Pomeranchuk}}
We consider a spinless Fermi liquid with an initially circular Fermi surface\cite{Pines}. 
Interactions between quasi-particles can be written in terms of an effective action of the form,
\begin{eqnarray}
  S_{\rm int} &=&
   \frac{N(0)}{2}
   \sum_{S,T}\int d^2xd^2x'dt\;
  \delta n_S({\bf x}) U_{S,T}({\bf x-x'})
  \delta n_T({\bf x'})\nonumber \\
   &+& \frac{\gamma N(0)}{4!}\sum_{S}\int d^2x dt\;\left(\delta n_S({\bf x})\right)^4.
\label{Sint}
\end{eqnarray}
Here, $N(0)$ is the density of states at the Fermi surface,  $S$ and $T$ label patches defined by coarse graining the Fermi surface, and the density fluctuations of quasi-particles in each patch is given by $\delta n_S({\bf x})=n_S({\bf x})-n^0_S({\bf x})$. The kernel $U_{S,T}({\bf x}-{\bf x}')$ is therefore the particle-hole pair interaction between the patches $S$ and $T$ and can be split into diagonal and off-diagonal components in the form, 
\begin{equation}
U_{S,T}({\bf x}-{\bf x}')= N(0) \delta(x-x')\delta_{S,T} + F_{S-T}(x-x').
\end{equation}
The quartic term in eq. \ref{Sint} could receive contributions from different processes, especially four body interactions in the fermion language. In ref. \onlinecite{BL}, a term of this class was obtained as a consequence 
of the curvature of the fermionic dispersion relation around the Fermi surface. While the quadratic term in eq. \ref{Sint} is marginal in the renormalization group sense\cite{FCN}, the quartic term is irrelevant. That means that in the isotropic phase, where the quadratic part is positive definite, the quartic term will not influence the asymptotic correlation functions (besides renormalization of the parameters of the model), obtaining the usual correlations associated with Landau Fermi liquids. However, near a Pomeranchuk instability, the quartic term is responsible for the stabilization of the anisotropic phases discussed in this article\cite{BL,Lawler}. 

Since the density fluctuation is  periodic around the Fermi surface, we can Fourier expand it as,
\begin{equation}
\delta n_S= \sqrt{\frac{2}{N}}\sum_{\ell=0}^{N/2} \rho_{\ell}\;\cos(\ell[\theta_S-\chi_\ell]),
\label{nS}
\end{equation}
and introduce the Fermi liquid parameters by means of
\begin{equation}
  F_{S-T} = \frac{1}{N} F_0 +
    \frac{2}{N}\sum_{\ell>0}F_\ell\cos\ell\left(\theta_S-\theta_T\right),
\label{FST}
\end{equation}
where $N$ is the number of patches covering the Fermi surface. Of course, at the end of the calculations, 
we  take the limit\cite{Marston,FCN} $N\to \infty$, keeping the density finite.

Since we are interested in the nematic/hexatic instabilities we will set, as a definition of our model,  $F_\ell=0$ for all $\ell\ne 2,6$. 
Essentially, we are assuming that all the other modes are stable. Integration over all the stable modes only renormalize the parameters since, very near criticality, they will only contribute with irrelevant operators. Notice that the quantities $(\rho_2,\chi_2)$ and $(\rho_6,\chi_6)$, in the expansion of the density fluctuation in eq. (\ref{nS}), are the nematic and hexatic order parameters introduced in eq. (\ref{orderparameter}) on general grounds.

Therefore, in order to write the model just in terms of our order parameters, we replace eqs. (\ref{nS}) and  (\ref{FST}) into the action (\ref{Sint}) finding, after summing up the contribution of different patches,  
that the effective action in the homogeneous limit has the same form given in eq. (\ref{Frho}), $S_{\rm int}\sim F$, with the identifications $\lambda_2=1+F_2$, $\lambda_6=1+F_6$, $\mu=2\gamma$ and $\beta=\gamma/3$.
With this values, the analysis of the phase diagram corresponds to that displayed in fig. \ref{figbp}, since $\mu>\gamma$ and $\beta<\beta_c$.

The values $\lambda_2=1+F_2<0$ and $\lambda_6=1+F_6<0$ correspond to  Pomeranchuk instabilities in the nematic and hexatic channels, respectively. They lead to deformations of the Fermi surface, since $k_F=k_0+ \delta n_S$ or, in terms of our model,
\begin{equation}
k_F(\theta_S)=k_0+ \rho_2 \cos(2\theta_S)-\rho_6\cos(6\theta_S),
\label{kF}
\end{equation}
where we have used the mean field solution for the angles, $\chi_6= \chi_2+(2n+1)\pi/6$, with $n=0,1,2$, and considered $\chi_2=0$, without loosing generality. The form of the Fermi surface for each phase is displayed in fig. \ref{figbp}. 

In the hexatic phase, we find,
\begin{equation}
\rho_2=0
\makebox[.5in]{,}
\rho_6=\sqrt{|1+F_6|/\gamma}.
\end{equation}
On the other hand, in the nematic phase,  we find  
\begin{equation}
\rho_2=\sqrt{|1+F_2|/\gamma}
\makebox[.5in]{,}
\rho_6\sim\frac{1}{3\gamma^{1/2}}\;\;\frac{|1+F_2|^{3/2}}{|1+F_6|+ 2|1+F_2|}.
\label{rho6}
\end{equation}
Therefore, all along the line of nematic criticality in fig. \ref{figbp}, where $1+F_2\sim 0$ but $1+F_6> 0$, the initially circular Fermi surface is deformed into an ellipse. Higher order harmonics tend to zero with $O(1+F_2)^{3/2}$. However, very close to the bicritical point, where $1+F_6\sim 0$, the hexatic harmonic in  (\ref{rho6}) becomes important and $\rho_6\sim 1/6\; \rho_2$. At this point, we are very near the first order phase transition, where the change form nematic to hexatic is discontinuous.  The energy difference between the
two local minima can be estimated as $\Delta{\cal F}\sim (3|1+F_2|-4|1+F_6|)/\gamma$.

\section{Collective modes \label{modes}}
The phase diagram of the model is completely determined by symmetry.  
However, the dynamics of each phase is dictated by quantum mechanics and it is not possible 
to deduce it by only using symmetry considerations. 

In order to study fluctuations around mean field, we will consider the complete action of the system including the dynamical part,  
\begin{equation}
S=S_{d}+S_{int},
\label{Scomplete}
\end{equation}
where $S_{int}$ is given by eq. {\ref{Sint}}. We will  choose $S_{d}$ as the usual dynamical term of the  Landau theory of Fermi liquids in the colitionless regime\cite{Pines},
\begin{equation}
S_{d}= \frac{1}{2}\sum_S\int d^2xdt\; \delta n_S \left(\frac{\partial_t}{\vec v_S\cdot\vec\nabla}\right)\delta n_S,
\label{Sd}
\end{equation}
where $\vec v_S$ is the Fermi velocity in the patch $S$ of the coarse grained Fermi surface. While this is a nonlocal action, sometimes it is useful to write a local version, introducing chiral bosons $\phi_S(x)$ defined by, 
\begin{equation}
\delta n_S(x)=N(0)\;\vec v_S\cdot \vec\nabla\phi_S(x).
\end{equation}
With this choice, the dynamical part of the action acquires the local form, 
\begin{equation}
S_{d}= \frac{N(0)}{2}\sum_S\int d^2xdt\; \partial_t\phi_S\delta n_S.
\label{Sdlocal}
\end{equation}
This is the usual expression in the context of higher dimensional bosonization\cite{FCN,Marston,BL,Lawler}, where the relevant degrees of freedom are $\phi_S$. In order to study collective modes, both the local and non-local versions of the action are equally convenient. However, the second local form in eq. (\ref{Sdlocal}) is more appropriate for the calculation of one particle properties.

We parametrized the fluctuations in the following form, 
\begin{equation}
\delta n_S(q)=\delta n_S^{\rm cl}+\varphi_S(q),
\label{fluctuation}
\end{equation}
where $\delta n_S^{\rm cl}$ is the {\em classical} mean field solution in each stable phase, evaluated in the previous section,  and $\varphi_S(q)$ is a small density fluctuation around this solution. 

In the isotropic phase, the non-quadratic terms can be ignored and, following the techniques developed in ref. \onlinecite{Lawler}, we find the following diagonal effective action for collective excitations,
\begin{equation}
S=\frac{1}{2}\sum_{j=1,2\; ; \eta=\pm } \int \frac{d^2qd\omega}{(2\pi)^3}\;
\xi^\eta_j\; M^\eta_j \xi^\eta_j\ ,  
\label{SM}
\end{equation}
\begin{equation}
\left( 
\begin{array}{c}
\xi^\pm_1 \\
\xi^\pm_2
\end{array}
\right)
=
{\bf A}^\pm(s)
\left( 
\begin{array}{c}
\varphi_2\pm \varphi_2^* \\
\varphi_6\pm \varphi_6^*
\end{array}
\right),
\end{equation}
where $\varphi_\ell$ is the Fourier transform of $\varphi_S$  and 
${\bf A}^\pm(s)$ is the dynamical matrix that diagonalizes the effective action. As usual, we use the notation $s=\omega/v_F q$.

The explicit expressions for $M_j$ and $A(s)$ depend on the regime we want to study. 
Along the two second order critical lines, except near the bicritical point,   we find ${\bf A}^\pm(s)= I + O(s)$. This means that, at leading order in $s$, the nematic and the hexatic modes are decoupled in this regime.  The kernel $M_j^\pm$ reads,  
\begin{eqnarray}
M^+_{j}&=&-1-F_j+2is - (-1)^{j} \frac{1-(F_6-F_2)}{F_6-F_2}\;s^2  \label{M+},\\
M^-_{j}&=&-1-F_j+2j\;s^2.
\label{M-}
\end{eqnarray}
From the last term in eq. (\ref{M+}), we see that the stable mode affects the critical mode only at next to leading order in $s$. This confirms our assumption that the stable modes can be integrated out, without modifying the asymptotic properties of the critical theory.

At the critical hexatic line we have $1+F_6\sim \kappa q^2$, where $\kappa$ is the range of the potential considered in the hexatic channel. Then, the  collective modes in this case are $\omega\sim i \kappa v_F q^3$ and  $\omega\sim \sqrt{\kappa} v_F q^2$, characterizing the critical hexatic phase,   and the other two stable nematic modes are the usual linear ones in a Fermi liquid, $\omega\sim  i (1+F_2) v_F q$ and  $\omega \sim \sqrt{1+F_2} v_F q$. This result is expected as it parallels the well known result for nematic fluctuations\cite{Lawler}. 

However, if the two order parameters become critical, the structure of the effective action changes substantially. Very near the bicritical point, where $F_2\sim F_6$, we find,
\begin{eqnarray}
M^+_j&=&-\kappa q^2 +(1+ (-1)^j) 2is+(-1)^j 4s^2, \\
M^-_j&=&-\kappa q^2 +4(2+(-1)^j \sqrt{2})s^2.
\end{eqnarray}
At this point we still have a cubic dissipative mode while the other three modes are stable quadratic ones.  
Then, the critical theory very near the bicritical point is also characterized by a dynamical exponent $z=3$. However, in this case it is impossible to distinguish between nematic and hexatic fluctuations since the matrix $A(s)$ strongly couples these modes.

In the symmetry broken phases, the order parameter (nematic or hexatic) picks up an orientation and 
the collective modes will necessary be anisotropic. We expect that the Goldstone modes, related to angular fluctuations, will dominate the low energy dynamics. Near criticality we obtain for each broken phase, 
\begin{eqnarray}
\lefteqn{
S_\ell=\frac{1}{2}\int\frac{d^2qd\omega}{(2\pi)^3}\times \nonumber }\\ 
&&\times\big(2|1+F_\ell(0)|+2is \cos^2(\ell(\phi-\chi_\ell) \big) \xi_1(q,\omega)\xi_1(-q,-\omega) \nonumber \\ 
&&+  \big(2is \sin^2\ell(\phi-\chi_\ell)-\kappa_\ell q^2\big) \xi_2(q,\omega)\xi_2(-q,-\omega),
\label{S6}
\end{eqnarray}
where $\ell=2,6$ and the angle $\phi$ is defined as $\vec q/q=(\cos\phi,\sin\phi)$.

The last term of this action indicates that the theory retains its critical character deep inside the anisotropic phases due to the $z=3$ Goldstone modes $\xi_2$. These modes exist for almost all  momenta, except for the special directions $\phi-\chi_\ell=n \pi/\ell$ with $n=0,..,2\ell-1$, where we have stable linear propagation. The two symmetry broken phases are separated by a discontinuity since $\chi_2-\chi_6=(2n+1)\pi/6$, with $n=0,1,2$.

\section{Summary and Conclusions \label{conclusions}}

In this paper we have characterized the isotropic/nematic/hexatic quantum phase transition 
in the vicinity of Pomeranchuk instabilities of a Fermi liquid.  

We have shown that the static effective action, or free energy, is completely determined by symmetry. The phase diagram, shown in fig.\ \ref{figbp},  contains two second order lines, corresponding to the isotropic/nematic and isotropic/hexatic phase transitions. Both continuous transitions meet together at a bicritical point where a first order nematic/hexatic transition emerges. It is interesting to note that the phases of the complex nematic and hexatic order parameters are not independent. Instead, they are coupled in a rotationally invariant way. The associated principal axes cannot be aligned, in fact, the possible values for $\chi_2-\chi_6$ are $\pi/6, \pi/2\; \mbox{or}\; 5/6\;\pi$. 

Since the presence of other two-body interactions, represented by stable Landau parameters, renormalizes the theory, the phase diagram  in fig. \ref{figtp} cannot be discarded. However, assuming that higher harmonics in the interactions are negligible, the phase diagram in fig. \ref{figbp} is more plausible.   

Quantum fluctuations have been computed in a region where  $F_2$ and $F_6$ are very near the Pomeranchuk instability. 
The important point is that on the whole critical region, even at the bicritical point where fluctuations in both channels cannot be decoupled, the theory is governed by a dynamical exponent $z=3$. This result validates our mean field treatment, since the effective dimensionality is above the upper critical dimension. In fact, a $z=3$ critical theory seems to be the fate of any spontaneously broken rotational symmetry in two dimensions, independently of the residual symmetry displayed by the ordered phase.   

On the broken symmetry side of the transition the theory continues to be critical due to  dissipative cubic Goldstone modes. These modes are proportional to the ``range''  $\kappa$ of the non-local Landau parameters. Therefore, even though the ordered phases exist for local interactions, they have zero stiffness, making them pathological. In this sense, although the non-locality of the two-body coupling is irrelevant in the renormalization group sense, it must be taken into account to correctly compute fluctuations. Near the bicritical point, both ordered phases are separated by a discontinuous transition. Thus, very near this region, the dynamics is related with metastability.
It is also important to underline that the presence of dissipative cubic modes profoundly modifies the asymptotic behavior of fermions, when compared with the usual one for Fermi liquids\cite{Lawler}.

We believe that this general analysis about the competition between Pomeranchuk instabilities will help to improve our understanding of the very complex phase diagram of real Hall liquids as well as other strongly correlated fermionic systems.

\begin{acknowledgments}
We would like to thank Eduardo Fradkin and Daniel A.\ Stariolo  for many interesting discussions and comments.
DGB would like to acknowledge the Department of Physics of the National University of La Plata, Argentina, for their kind hospitality during part of this work. 
The Brazilian agencies Conselho Nacional de Desenvolvimento Cient\'{\i}fico e Tecnol\'{o}gico (CNPq),  and the Funda{\c {c}}{\~{a}}o de Amparo {\`{a}} Pesquisa do Estado do Rio de Janeiro (FAPERJ) are acknowledged for the financial support.
\end{acknowledgments}

\end{document}